\documentstyle[12pt,epsf,epsfig]{article} 
\begin{document}

\vglue 1.0 cm

\begin{flushright}
  {TSU HEPI 2000-04 } 
\end{flushright}         

\vglue 3.5cm 

\begin{center}
{\Large \bf Unusual Bound States in QFT Models }
\medskip

\vglue 1.0cm
{\bf A.\ A.\ Khelashvili\footnote {e-mail:temo@hepi.edu.ge}}
\medskip

{\it High Energy Physics Institute, Tbilisi State University, 380086,
 Tbilisi, Georgia}
% \end{center}

\vglue 2.5cm
 
{\bf { ABSTRACT }}
\vglue 2.0cm

\parbox{6.4in}{\leftskip=1.0pc  

 Homogeneous Bethe-Salpeter equation for simplest Wick-Cutkosky model is
studied in the case when the mass of the two-body system is more then the
sum of constituent particles masses. It is shown that there is always a
small attraction between the like-sigh charged particles as a pure
relativistic effect. If the coupling constant exceeds some critical values
there arise discrete levels.The situation here is analogous to the
so-called "abnormal" solutions.The signature of the norm of these discrete
states coincides with the "time-parity".The states with the negative norms
can be excluded from the physical sector-the one-time (quasipotential)
wave-function corresponding to them vanishes identically.However the
positive norm states survive and contribute to the total Green function
(and the S-matrix) with the proper sign.  }
                                          
\end{center}
\vglue 2.0cm
  
\newpage

\medskip

\section{Introduction}

The main aim of this article is to demonstrate that equations of the
relativistic quantum theory may  have solutions which have no
nonrelativstic analogue. Some of such solutions are
well-known, e.g.\ the so-called abnormal solutions. Nakanishi wrote
about the latters: "The non-relativistic common sense does not
necessarily remain valid in the relativistic quantum field
theory'' \cite{1}.

The main ideas of this article were published as a preprint of
Tbilisi Mathematical Institute in collaboration with A.\
Tavkhelidze and L.\ Vachnadze \cite{2}. 

At first the interest to the problem was initiated by the
observed narrow peaks around $1.5MeV$ in the $e^+e^-$ distribution
in the experiments of heavy ion collisions \cite{3} . These
resonances are known as the GSI resonances. They are situated above
the $2m_e$ threshold and it naturally triggered interest to studies
of bound states in continuum. In the papers of the Serpukhov and
other groups \cite{5} the single-photon exchange quasipotential
equations in QED were studied and, relying on the numeric methods,
large number of peaks were reported.

The quasipotential equation can be obtained from the
Bethe-Salpeter equation after certain approximation. So it was
natural to try to cope with the same problem within the framework of
the ladder approximation of the Bethe-Salpeter equation itself.
It was particularly interesting because there was no such
observation made on the earlier stages of studies of the
Bethe-Salpeter (BS) equation (for a review see \cite{1}). However,
we think that nobody has searched for bound states in the continuum
there. It is worth noting that when the total mass $M>2m$ the
traditional method --- Wick's rotation is, in general,
inadmissible.  So the problem must be considered in the Minkowski
space. Structure of the ladder BS equation for this case was
studied by G\"unter \cite{7} and it was demonstrated that Wick's
rotation is inadmissible. In the recent years interest to the BS
equations in the Minkowski space has increased \cite{8},\cite{9}.
Below, employing standard methods, we will reduce the BS equation
to the system of the one-dimensional equations and study unusual
solutions in the continuous spectrum. Note that though further
experiments have not confirmed existence of the GSI resonances the
possibility of bound states embedded in the continuum is
interesting on its own.

\section{Bethe-Salpeter equation in the Ladder approximation and
its reduction}

We will consider the BS equation in the ladder approximation for a
simple model \cite{6},\cite{10}  of two equal mass ($m$) scalar
particles exchanging massless scalar meson, the so called
Wick-Cutkosky model with the interaction Lagrangian:
$$ L_{int}=g_1\phi_1^*\phi_2 A+g_2\phi_2^*\phi_1 A. $$
Standard equation has the form (in the rest frame):
\begin{equation}
\psi(p)=-\frac{\lambda}{\pi^2i}\int d^4q\frac{\psi(q)}
{\left[\left(p-q\right)^2+io\right]
\left[\left(\frac{P}{2}-q\right)^2-m^2\right]
\left[\left(\frac{P}{2}+q\right)^2-m^2\right]}.
\label{1}
\end{equation}
Here $P=p_1+p_2$ is the total 4-momentum, $p=\frac{1}{2}(p_1-p_2)$
and $q$ are relative 4-momenta, $\lambda=\frac{g_1g_2}{(4\pi)^2}$
is the product of appropriate coupling constants and has
dimension of the mass squared. In the rest frame where $\vec P=0$
and $\vec p_1=-\vec p_2=\vec p$ we will use the following notation
for the total energy (mass) $P_0=E=M=2W$.

Usually the Wick-Cutkosky model is considered as an auxiliary
(fictitious) one. But in some reasonable approximation more
realistic models reduce to the Wick-Cutkosky one. Indeed, in the
single-photon (gluon) approximation in QED (QCD) for fermion and
antifermion bound state BS amplitude we have an equation:
$$
\left[m-\gamma(\frac{P}{2}+p)\right]\psi(p)
\left[m+\gamma(\frac{P}{2}-p)\right]=
=\frac{\lambda_F}{\pi^2i}\int
d^4q\frac{\gamma_\mu\psi(q)\gamma_\mu}{(p-q)^2+io}
$$
where $\lambda_F\equiv\left(\frac{g}{4\pi}\right)^2$ and the
Feynman gauge is used.

Usually a new function
$$\Psi(p)=
\left[m-\gamma(\frac{P}{2}+p)\right]\psi(p)
\left[m+\gamma(\frac{P}{2}-p)\right]
$$
is introduced \cite{1} and the decomposition into the Dirac
structures is considered:
$$
\Psi=\Psi^S-\gamma_5\Psi^P+\gamma^\mu\Psi^V_\mu-\gamma^\mu\gamma_5
\Psi^A_\mu-\frac{1}{2}[\gamma^\mu,\gamma^\nu]\Psi_{\mu\nu}^T
$$

Further one can obtain a chain of coupled equations for Dirac
structures which can be  considerably simplified by taking
binding energies small (i.e.\ for weakly coupled system).
All the components of $\Psi$ that survive in that limit of small
binding energies ($\Psi^P,\ \Psi^A_\mu,\ \Psi^V$) satisfy the same
equations:
$$
\Psi^i(p)=
=-\frac{4\lambda_Fm^2}{\pi^2i}\int
d^4q\frac{\Psi^i(q)}{\left[(p-q)^2+io\right]
\left[\left(\frac{P}{2}-q\right)^2-m^2\right]
\left[\left(\frac{P}{2}+q\right)^2-m^2\right]} $$
which coincides with the equations of Wick-Cutkosky model if we
set $4\lambda_Fm^2=\lambda$ or $\lambda\pi=m^2\alpha$    where
$\alpha=e^2/4\pi$. Therefore for weekly coupled systems in
single-photon (gluon) exchange approximation in QED (QCD) we have
in fact to deal with the equations of Wick-Cutkosky model.

It is well known \cite{6},\cite{10} that in the case $W<m$ and
$\lambda>0$ (which corresponds to attraction) with small $\lambda$
eq.(\ref{1}) has a discrete spectrum in accordance with the Balmer
series for the hydrogen atom.
Now our task is to show whether this equation has any discrete
solutions when $W>m$ i.e.\ $M>2m$.

It is convenient to transform the integral BS equation into a
differential one in momentum space by using the identity:
\begin{equation}
\left[\frac{\partial^2}{\partial p_0^2}-\frac{\partial^2}{\partial
\vec p^2} \right]\frac{1}{(p-q)^2+io} = -4\pi^2i\delta^{(4)}(p-q).
 \label{2}
\end{equation}
We obtain:
\begin{equation}
\left[\frac{\partial^2}{\partial p_0^2}-\frac{\partial^2}{\partial
\vec p^2} \right]\Psi(p) = \frac{4\lambda_F\Psi(p)}
{(p_0^2-\vec p^2+W^2-m^2)^2-4p_0^2W^2)}
\label{3}
\end{equation}

This equation is mathematically equivalent to eq.(\ref{1}) if
$\Psi(p)$ satisfies the following boundary conditions \cite{14}:
$p^2\Psi(p)$ remains finite for large values of $p$ and $\Psi(p)$
must be finite when $p=0$.

We think that like in the Coulomb problem in our problem too it is
possible to separate variables due to the hidden additional
symmetry of the Coulombic interaction \cite{13}. For earlier
attempts of variable separation see \cite{12},\cite{13},\cite{14}.

First of all let us single out the spherical angles in a standard
way:
$$\Psi(p)=\Phi(p_0,p_s)Y_{lm}(\theta,\phi)$$
where the spherical coordinates are defined as follows:
$$p_1=p_s\sin(\theta)\cos(\phi),\ \ \
p_2=p_s\sin(\theta)\sin(\phi),\ \ \ p_3=p_s\cos(\theta), \ \ \
p_s=\pm |\vec p|.$$
Then eq.(\ref{3}) is reduced to the partial differential equation
with two variables:
\begin{equation}
\left[\frac{\partial^2}{\partial p_0^2}-\frac{\partial^2}{\partial
p_s^2} \right](p_s\Phi(p) = \frac{4\lambda_F(p_s\Phi(p)}
{(p_0^2-p_s^2+W^2-m^2)^2-4p_0^2W^2)}
\label{4}
\end{equation}

Boundary conditions for $\Phi(p_0,p_s)$ can be rewritten as
$$p_s\Phi(p_0,p_s)\biggr|_{p_s\to 0}\to 0\ \ \ \ \
p_s\Phi(p_0,p_s)\biggr|_{p_0\to 0} < \infty$$
\begin{equation}
\frac{p^2}{p_s}(p_s\Phi(p_0,p_s))\biggr|_{p\to 0}<\infty
\label{5}
\end{equation}

To investigate the energy spectrum in the region$W>m$ let us use
the method of bipolar transformations
\cite{11},\cite{13},\cite{14}. Peculiarity of our case is that the
whole $(p_0,p_s)$-plane can be covered only by the means of
several such transformations \cite{14}:
\begin{eqnarray}
{\it (1)} & p_0=\frac{k \sinh\alpha_1}{\cosh\alpha_1+\cosh\beta_1},
\ \ \ p_s=\frac{k\sinh\beta_1}{\cosh\alpha_1+\cosh\beta_1} &
\nonumber \\
{\it (2)} & p_0=\frac{k \sinh\alpha_2}{\cosh\alpha_2-\cosh\beta_2},
\ \ \ p_s=\frac{k\sinh\beta_2}{\cosh\alpha_2-\cosh\beta_2} &
|\alpha_2|\neq|\beta_2|
\nonumber \\
{\it (3)} & p_0=\frac{k \cosh\alpha_3}{\sinh\alpha_3+\sinh\beta_3},
\ \ \ p_s=\frac{k\cosh\beta_3}{\sinh\alpha_3+\sinh\beta_3} &
\alpha_3\neq-\beta_3
\nonumber \\
{\it (4)} & p_0=\frac{k \cosh\alpha_4}{\sinh\alpha_4+\sinh\beta_4},
\ \ \ p_s=-\frac{k\cosh\beta_4}{\sinh\alpha_4+\sinh\beta_4} &
\alpha_4=-\beta_4.
\nonumber
\end{eqnarray}

The hyperbolic angles in all these transformations run over the
whole infinite intervals:
$$-\infty<\alpha_i,\ \beta_i<\infty$$
while the parameter $k$ is determined by the condition of
separabilty in eq.(\ref{3}) and equals to $(W^2-m^2)^{1/2}$.

Action of these transformations is shown in Fig.1. 
After performing these
transformations variables in eq.(\ref{4})
separate:
$$
p_s\Phi(p_0,p_s)=f(\alpha_i)g(\beta_i)\ \ \ \ \ \ \ i=1,2,3,4
$$
and we obtain the following equations for each of the functions
$f$ and $g$:

\begin{description}
\item{(a)} For transformations $(1)-(2)$ (the shaded regions in
 Fig.1):
 \begin{equation}
 \left[\frac{d^2}{d\beta^2}+h_1-\frac{l(l+1)}{\sinh^2\beta}
  \right] g(\beta)=0,
\label{6} \end{equation}
\begin{equation}
 \left[\frac{d^2}{d\alpha^2}+h_1-\frac{\lambda/m^2}{a-\cosh^2\alpha}
  \right]f(\alpha)=0;
   \label{7}
 \end{equation}

\item{(b)} For transformations $(3)-(4)$ (the unshaded regions in
 Fig.1):
 \begin{equation}
 \left[\frac{d^2}{d\beta^2}+h_2+\frac{l(l+1)}{\cosh^2\beta}
  \right] g(\beta)=0,
\label{8} \end{equation}
\begin{equation}
 \left[\frac{d^2}{d\alpha^2}+h_2-\frac{\lambda/m^2}{a+\sinh^2\alpha}
  \right]f(\alpha)=0.
   \label{9}
 \end{equation}
\end{description}
In these equations $$ a\equiv\frac{W^2}{m^2}>1  $$ and $h_1, h_2$
are the separation parameters (new quantum numbers). The fact of
separability is provided by the symmetry of relativistic problem in
the 4-dimensional world \cite{12}.

Equations obtained above describe one-dimensional motions in the
space of hyperbolic angles. Equations for $g(\beta)$ are kinematic
ones defining the values of separation parameters $h_{1,2}$.
 Equations for $f(\alpha)$
contain dynamical quantities $(\lambda,a)$ and hence must allow to find
the energy spectrum.

Boundary conditions for $g(\beta)$ and $f(\alpha)$ follow
immediately from (\ref{5}) and are the same in different
regions of the $(p_0,p_s)$-plane
\begin{eqnarray}
& g(0)=0 & g(\pm\infty)<\infty
\label{10} \\
& f(0)<\infty & f(\pm\infty)=0
\label{11}
\end{eqnarray}

Note that the boundaries of different regions in $(p_0,p_s)$-plane
can be approached by taking limits $\alpha,\beta\to \pm\infty$
in the relevant transformations $(1)-(4)$.

\section{Solutions of the one-dimensional equations}

First consider the pair (\ref{6})-(\ref{7}). Eq.(\ref{6}) with
the boundary conditions (\ref{10}) has the following solution:
\begin{equation}
g(\beta)=g_l(\beta)\sim \Biggl\{
\sinh^{l+1}\beta
{\left[
\frac{1}{\sinh\beta}
\frac{d}{d\beta}
\right]^{l+1}
\cosh(\sqrt{h_1}\beta),
\ \ \ \ h_1 > 0 }
\atop
{0, \qquad\qquad h_1<0}
\label{12}
\end{equation}

In general, the second equation of this pair can give the relation
between energy $W$ and the parameter $h_1$
$$ W=\rho_n(h_1), \ \ \ \ \ \\ n=0,1,2\ldots $$

Despite the fact that $n$ can take discrete values, energy
is not quantized because $h_1$ remains to be an artbitrary
positive number.

Things are  essentially different in the unshaded regions
$(3)-(4)$. Eq.(\ref{8}) has solutions with positive and negative
signs of $h_2$. The positive $h_2$ are not resulted while the
negative ones are quanttizes for $l>0$.
$$
h_2=-(l-N)^2, \ \ \ \ \ \ N<l.
 $$
 where $N$ takes odd integer values.

 Next consider eq.(\ref{9}) in more detail. Expression
 $$
 V(\alpha)=\frac{\lambda/m^2}{a+\sinh^2\alpha}
 $$
 plays the role of "potential".

 For positive $\lambda$ (i.e.\ attraction) $V(\alpha)$ is the
 positive definite and convex function of $\alpha$ (Fig.2).
 Obviously we can have only continuous spectrum in this case witth
 posititve $h_2$.

When we take $\lambda$ negative (repulsion) $V(\alpha)$ becomes
negative definitte and concave function of $\alpha$ (Fig.2).
Therefore in this case there appears principal possibility to have
discrete eigenvalues provided the depth of $V(\alpha)$ is high
enough for a given negative value of $h_2=-(l-N)^2$.

One can examine easily that continuity on the boundaries of shaded
and unshaded regions imposes no new constraints on separation
parameters $h_{1,2}$. So there is no discrete specttrum for
positive values of $\lambda$ when $W>m$.

Now we turn back to eqs.(\ref{8})-(\ref{9}) with $h_2<0$ and
$\lambda<0$. Eq.(\ref{8}) is a Heun's differential equation
\cite{16}. Its solution with $h_2=(l-N)^2$ and
$\lambda=-|\lambda|<0$ imposed by boundary and normalization
conditions leads to complicated transcendential equations. To
estimate qualitatively the coupling constant dependence of the
energy eigenvalues one can make use of tthe following approximate
expression
$$W^2\sim \frac{|\lambda|}{\nu(\nu+1)} \ \ \ \ \ \nu=1,2,\ldots$$
As $W^2>m^2$ it follows that
$$
\frac{\lambda^2}{m^2}\geq\nu(\nu+1)=2,6,12,\ldots
$$
i.e.\ discrete levels apeear when $|\lambda|$ exceeds some
critical values. It is worth noting that for some fixed values of
$\lambda^2/m^2$ there is finite number of discrete levels. For
example, when
$$
2<\frac{\lambda^2}{m^2}<6
$$
there is only a single level. And for
                                     $$
6<\frac{\lambda^2}{m^2}<12 $$
the number of levels equals $2$, etc.

If we take $|\lambda|$ in accordance with Balmer formula mentioned
above in the following form $\lambda\pi=m^2Z^2\alpha$, then the
above evaluation can be cast into
$$
\frac{Z^2\alpha}{\pi}>\nu(\nu+1)
                                $$
i.e.\ $Z_{min}>30$.

We want to note tthat eqs.(\ref{8})-(\ref{9}) were derived earlier
in \cite{13} but with erroneous sign of $\lambda$. Hence author
deduced bound states for positive values of $\lambda$.

\section{Normalization condition}

Let us now discuss the normalization condition for the BS
amplitude, which is nonvanishing only in the unshaded regions and
has the form:
\begin{equation}
\psi(p)\sim \frac{(\sinh\alpha+\sinh\beta)^3f(\alpha)g(\beta)}
{(W^2-m^2)^{3/2}\cosh^3\beta(W^2+m^2\sinh^2\alpha)}Y_{lm}(\theta,\phi)
\label{13}
\end{equation}

The role of normalization is to provide the correct relation
between the aplitude $\psi$ and four-point Green function. The
exact form of the normalization condititon for the model under
consideration looks like \cite{1}
\begin{equation}
  i\int d^4p\bar\psi(p) G_0^{-1}(p)\psi(p)=\varepsilon_B\lambda
\frac{dS_B(\lambda)}{d\lambda}
\label{14}
\end{equation}
where $S_B(\lambda)=4W^2_B(\lambda)$ and $\varepsilon_B$ is the so
called norm factor determining the sign of the pole contribution in
the total Green function at $S\sim S_B(\lambda)$
\begin{equation}
G\sim i\varepsilon_B\frac{\psi_B(p)\bar\psi_B(p)}{S-S_B}+\mbox{Reg}
\label{15}
\end{equation}

Integration in the normalization condition (\ref{14}) is carried
 over the unshaded region only.

 The spectral representation for $\psi(p)$ and $\bar\psi(p)$ that
 we are going to use has the form \cite{1}:
 \begin{equation}
 \psi(p_0,\vec p)=-\frac{1}{2\pi i}\int^{\infty}_{w_+(\vec p)}
 dq_0\frac{f(q_0,\vec p,W)}{p_0-q_0+io}+
 \frac{1}{2\pi i}\int_{-\infty}^{w_-(\vec p)}
 dq_0\frac{g(q_0,\vec p,W)}{p_0-q_0-io};
  \label{16}
 \end{equation}
 \begin{equation}
 \bar\psi(p_0,\vec p)=-\frac{1}{2\pi i}\int^{\infty}_{w_+(\vec p)}
 dq_0\frac{f^*(q_0,\vec p,W)}{p_0-q_0+io}+
 \frac{1}{2\pi i}\int_{-\infty}^{w_-(\vec p)}
 dq_0\frac{g^*(q_0,\vec p,W)}{p_0-q_0-io}
  \label{17}
 \end{equation}
 where $w_-(\vec p)=W-(\vec p^2+m^2)^{1/2},\ \ w_+(\vec
 p)=-w_-(\vec p)$ are the enndpoints where cuts in the complex
 $p_0$-plane extend from. As long as 3-momentum $\vec p$ satisfies
$0\leq p_s^2\leq W^2-m^2$ these two cuts overlap (Fig.3).

 The spectral reprezentations (\ref{16})-(\ref{17}) imply
\begin{equation}
\bar\psi(z)=-\psi^*(z^*), \ \ \ \ \ z=\Re p_0+i\Im p_0
  \label{18}
 \end{equation}

There are ``dangerous" regions in the normalization integral where
left-hand and right-hand cuts overlap. This overlapping happens
when
\begin{eqnarray}
&  p_0\in {-w_0,0} & \mbox{for right-hand cut}, \nonumber \\
&  p_0\in {0,w_0,0} & \mbox{for left-hand cut}. \nonumber
\end{eqnarray}
In these intervals $0\leq p_s^2\leq k^2=W^2-m^2$. In the
$(p_0,p_s)$-plane ``dangerous" points are located between the
profiles of functions $w_+(\vec p)$ and $w_-(\vec p)$ for
$p_s^2\leq k^2$. This region is drawn in Fig.4.

Note that the ``dangerous" points completely lie in the region
where our BS amplitude vanishes identically. Therefore, owing to
this specific behaviour of the BS amplitude for bound states with
$W>m$, intervals of $p_0$ where cuts overlap do not contribute to
the normalization integral.

It seems to us that this property is very important and can show
up in more interesting cases too. Specifically this fact may be
employed for confining kernels, because sometimes bound states
above threshold ($W>m$) do arise for such kernels. Example
considered above indicates that in similar cases the
$(p_0,p_s)$-plane may bear some structure --- namely the BS
amplitude may have support region in the $(p_0,p_s)$-plane.
So effectively we can take that cuts in the spectral representation
start from the origin ($p_0=0$) and, consequently, perform Wick
rotation without any complications. This is the case when Wick
rotation is admissible for dynamical reasons. As a result we
obtain:
\begin{equation}
i^2\int d^3pdp_4 \bar\psi(ip_4,\vec p) G_0^{-1}(ip_4,\vec
p)\psi(ip_4,\vec p)=\varepsilon\lambda\frac{dS_B(\lambda)}{d\lambda}
\label{19} \end{equation}
Now, according to (\ref{18}) and symmetry of eq.(\ref{3}) under the
relative time reflection $p_0\to -p_0$ we have:
\begin{equation}
\bar\psi(ip_4,\vec p)=\psi^*(-ip_4,\vec p)=-\Pi_B\psi^*(ip_4,\vec
p), \label{20} \end{equation}
where $\Pi_B$ denotes the so-called relative-time parity
(``$p_0$-parity") of the bound state.

Using (\ref{20}) in the normalization integral we get
$$
\int d^3pdp_4 \bar\psi(ip_4,\vec p) G_0^{-1}(ip_4,\vec
p)\psi(ip_4,\vec p)=\varepsilon\lambda\frac{dS_B(\lambda)}{d\lambda}
$$
As long as $G_0^{-1}(ip_4,\vec p)$ is positive definite in the
unshaded regions the full integral above is positive definite too.
Besides, from the explicit form of our solution we know that
$$
\lambda\frac{dS_B(\lambda)}{d\lambda}=
|\lambda|\frac{dS_B(\lambda)}{d|\lambda|}>0
$$
and therefore with appropriate choice of overall multiplicative
constant up to which the BS amplitude is determined the norm factor
$\varepsilon_B$ coinsides with ``$p_0$-parity" of the bound state
$$ \varepsilon_B=\Pi_B\ .$$
In other words the states with positive (negative) ``$p_0$-parity"
have the positive (negative) norm.

States with the negative norm can be be eliminated in from the
physical sector as the correspondinng ``one-time" quasipotential
wave functions \cite{17} vanish identically:
$$
\int^{\infty}_{-\infty} dp_0 \psi(p_0,\vec p)=0\ .
$$

On the other hand the positive norm states, i.e.\ the states with
positive ``$p_0$-parity" do survive and produce the pole
contributions to the total Green function (and S-matrix) with the
correct sign.

\section{Conclusions}

We have demonstrated that in the BS equation for the equal mass
scalar particles' repulsive ($\lambda<0$) interaction via scalar
massless particle exchange there emerges an effective attraction
above thershold. This attraction can lead to existence of bound
states for values of the coupling $\lambda$ larger then the
critical one:
$$\frac{|\lambda_c|}{m^2}>2\ . $$
In QED with the one-photon exchange this corresponds to
$$\frac{\alpha_c}{\pi}>2\ .$$

It is worth mentioning that attraction is present for {\it all}
values of $\lambda$ as a purely relativistic effect in the
kinematic region where the retardation effects can not be treated
by perturbative methods.

\newpage

\begin{center}
%%%%%%%%%%%%%%%%%%%%%%%%%
%  Figure captions
%%%%%%%%%%%%%%%%%%%%%%%%%
%
%%%% fig 1
%
% \vspace{-3cm}
\begin{figure}[t]
\begin{minipage}[b] {12cm}
\epsfig{file=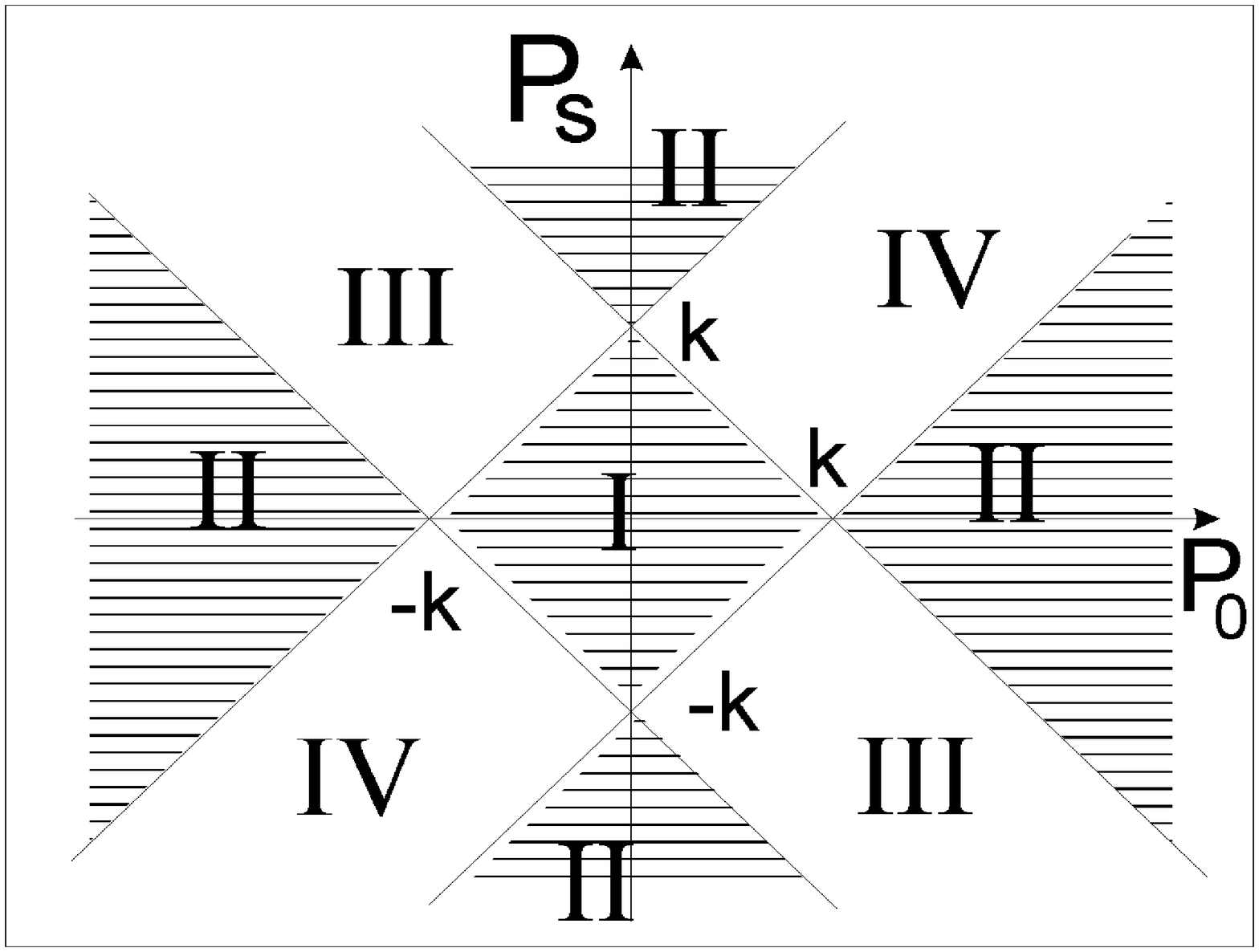,height=9cm,width=11cm}
\caption 
 { Regions in $p_0,\ p_s$ plane corresponding to bipolar  transformations
{\it (1-4)} }  
\end{minipage}
\end{figure}
%
%%%% fig 2
%
\begin{figure}[t]
\begin{minipage}[b]{12cm}
\epsfig{file=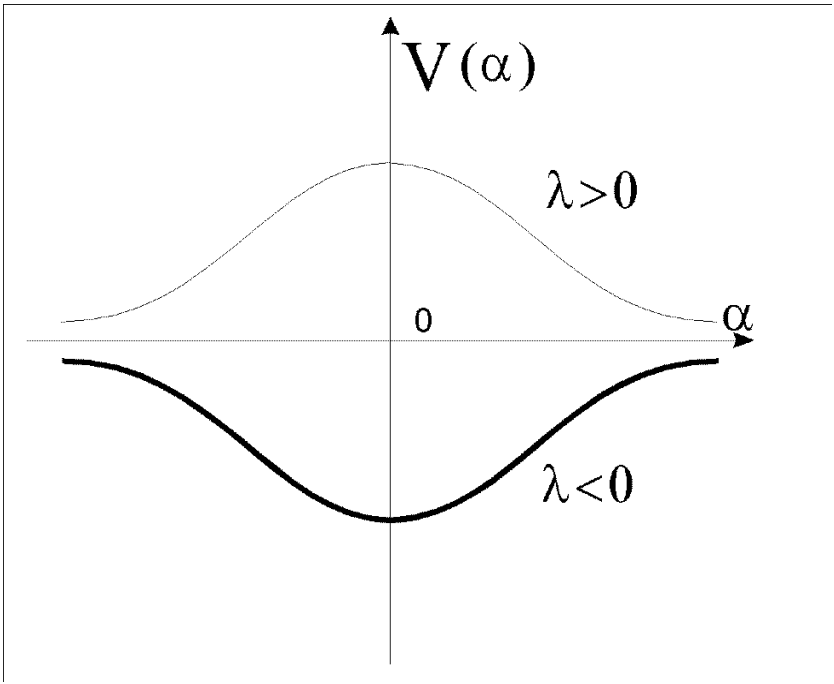,height=9cm,width=11cm}
\caption 
  { The ``potential'' $V(\alpha)$ for both signes of $\lambda$. } 
\end{minipage}
\end{figure}
%
%%%% fig 3
%
\begin{figure}[t]
\begin{minipage}[b]{12cm}
\epsfig{file=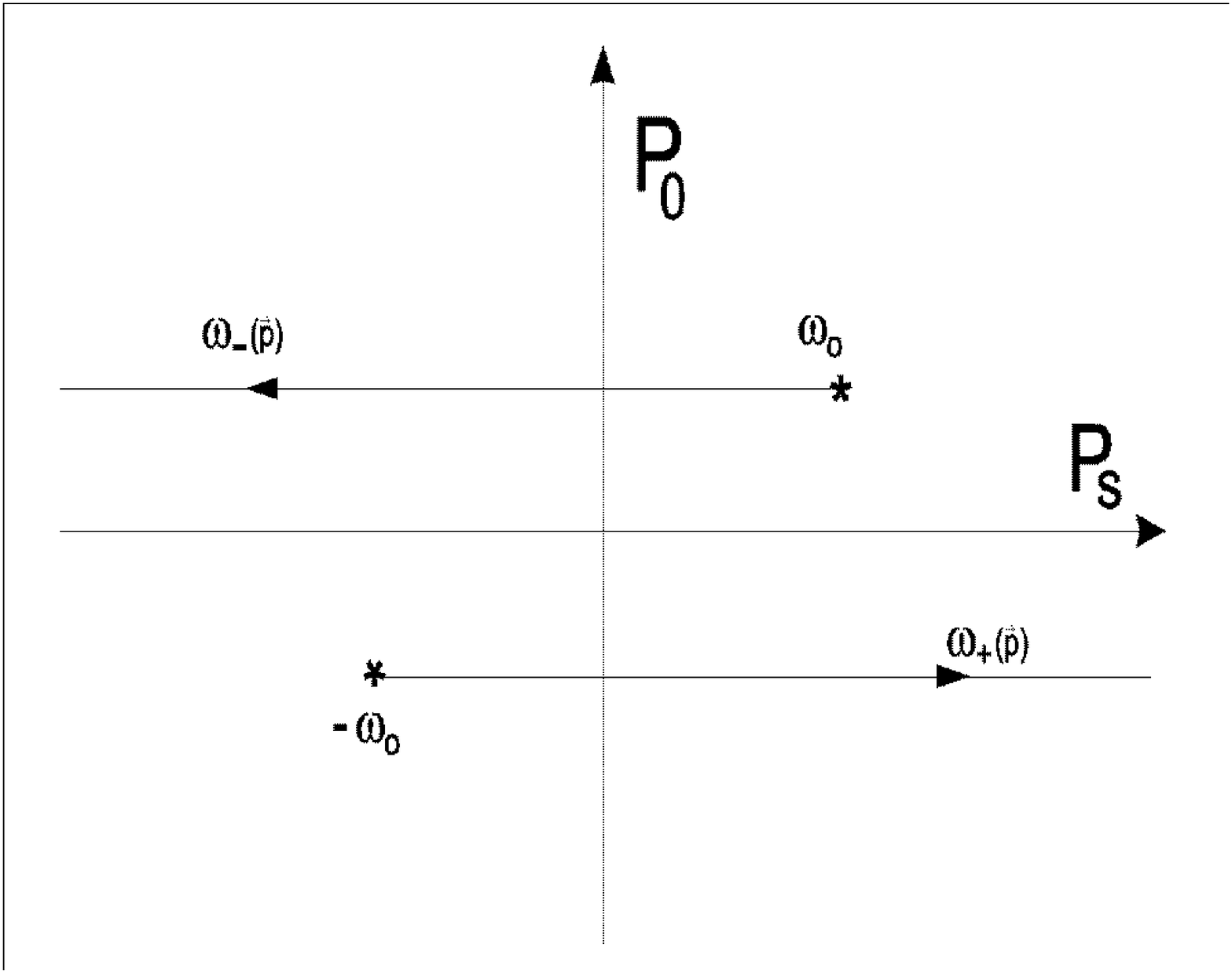,height=9cm,width=11cm}
\caption 
  { Complex $p_0$ plane. Here $\omega_0=[\omega_-(\vec p)_{max}=W-m>0$. }
\end{minipage}
\end{figure}
%
%%%% fig 4
%
\begin{figure}
\begin{minipage}[b]{12cm}
\epsfig{file=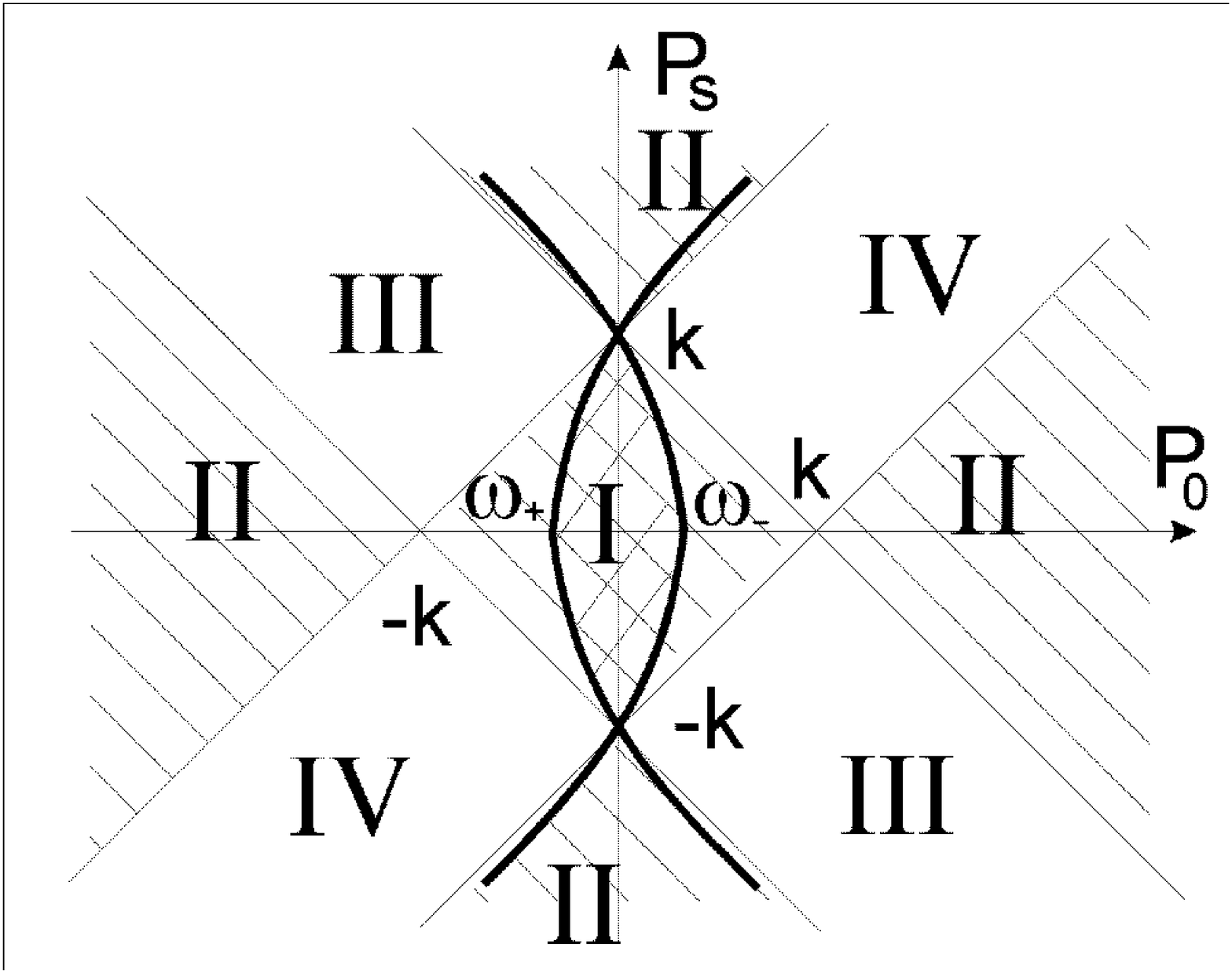,height=9cm,width=11cm}
\caption 
{  ``Dangerous'' region in the $p_0,\ p_s$ plane.} 
\end{minipage}
\end{figure}
\end{center}

\end{document}